\documentclass[reprint,aps,superscriptaddress,amsmath,longbibliography]{revtex4-2}
\usepackage{graphicx,txfonts,dcolumn,multirow,color}
\usepackage[colorlinks,linkcolor=blue,citecolor=blue,anchorcolor=blue,urlcolor=blue]{hyperref}

\begin{document}

\title{Explosive percolation in finite dimensions}

\author{Ming Li}
\email{lim@hfut.edu.cn}
\affiliation{School of Physics, Hefei University of Technology, Hefei, Anhui 230009, China}

\author{Junfeng Wang}
\affiliation{School of Physics, Hefei University of Technology, Hefei, Anhui 230009, China}

\author{Youjin Deng}
\email{yjdeng@ustc.edu.cn}
\affiliation{Hefei National Research Center for Physical Sciences at the Microscale, University of Science and Technology of China, Hefei 230026, China}
\affiliation{Department of Modern Physics, University of Science and Technology of China, Hefei, Anhui 230026, China}
\affiliation{Hefei National Laboratory, University of Science and Technology of China, Hefei, Anhui 230088, China}

\date{\today}

\begin{abstract}
Explosive percolation (EP) has received significant research attention due to its rich and anomalous phenomena near criticality. In our recent study [\href{https://doi.org/10.1103/physrevlett.130.147101}{Phys. Rev. Lett. 130, 147101 (2023)}], we demonstrated that the correct critical behaviors of EP in infinite dimensions (complete graph) can be accurately extracted using the event-based method, with finite-size scaling behaviors still described by the standard finite-size scaling theory. We perform an extensive simulation of EPs on hypercubic lattices ranging from dimensions $d=2$ to $6$, and find that the critical behaviors consistently obey the standard finite-size scaling theory. Consequently, we obtain a high-precision determination of the percolation thresholds and critical exponents, revealing that EPs governed by the product and sum rules belong to different universality classes. Remarkably, despite the mean of the dynamic pseudocritical point $\mathcal{T}_L$ deviating from the infinite-lattice criticality by a distance determined by the $d$-dependent correlation-length exponent, $\mathcal{T}_L$ follows a normal (Gaussian) distribution across all dimensions, with a standard deviation proportional to $1/\sqrt{V}$, where $V$ denotes the system volume. A theoretical argument associated with the central-limit theorem is further proposed to understand the probability distribution of $\mathcal{T}_L$. These findings offer a comprehensive understanding of critical behaviors in EPs across various dimensions, revealing a different dimension-dependence compared to standard bond percolation.
\end{abstract}

\maketitle

\section{Introduction}

Explosive percolation (EP) refers to the emergence of connected clusters in a modified bond percolation model, often realized through the Achlioptas process~\cite{Achlioptas2009}. Unlike standard bond percolation, where the inserted bonds are randomly and independently chosen from the system, the Achlioptas process selects bonds based on rules that can suppress the growth of large clusters. A typical one is the product rule. Starting from a lattice/graph with all bonds empty, at each time step two potential bonds are chosen randomly, and the one with the smaller size product of the associated clusters is eventually inserted. Another variant is the sum rule, which inserts the bond leading to a smaller total size of the associated clusters.

Compared to standard bond percolation, the onset of the percolation phase transition in EP can be significantly delayed. However, once the percolation threshold is surpassed, EP exhibits a remarkably sharp transition, resulting in the sudden formation of large-scale clusters. At the earlier days, this sharp transition was perceived as inducing a discontinuous transition~\cite{Achlioptas2009,Friedman2009,Ziff2009,Cho2009,Radicchi2009,Ziff2010,Radicchi2010,DSouza2010,Cho2011}. Subsequent studies, however, demonstrated that EP undergoes a continuous transition~\cite{Costa2010,Lee2011,Grassberger2011,Riordan2011,DSouza2015}. It has been theoretically proven that EP is always continuous unless subject to global dynamics~\cite{Riordan2011}, as seen with the rules used in Refs.~\cite{Araujo2010,Chen2011,Cho2013}. Despite its continuous nature, EP exhibits rich critical phenomena, such as the powder keg mechanism~\cite{Friedman2009}, bimodal distribution of the order parameter~\cite{Grassberger2011,Tian2012}, and non-self-averaging behavior~\cite{Riordan2012}. These phenomena are anomalous because they are typically associated with discontinuous phase transitions. Such behaviors have been observed in various systems, including regular lattices~\cite{Ziff2009,Ziff2010,Wu2022}, scale-free networks~\cite{Cho2009,Radicchi2009}, and growing networks~\cite{Yi2013,Oh2016,Oh2018,Oh2019,Oh2021}, as well as with different bond-insertion rules ~\cite{Friedman2009,Costa2010,DSouza2010,Nagler2011,Riordan2012,Costa2015}. Several scaling formalisms have also been proposed to explain these anomalous finite-size behaviors~\cite{Cho2010,Bastas2011,Li2012,Costa2014,Sabbir2018}.

Our recent study~\cite{Li2023} has shown that in infinite dimensions, i.e., on complete graphs, EP obeys the standard finite-size scaling (FSS) theory when using an event-based ensemble. This approach allows for highly precise determination of the percolation threshold and critical exponents. As a result, an understanding is provided that the bimodal distribution of the order parameter at the infinite-lattice critical point results from a mixed effect over a wide range that exceeds the scaling window centered on the dynamic pseudo-critical point. Moreover, on the basis of the crossover FSS theory~\cite{Li2024}, the multiple scalings as observed in the bimodal distribution can be derived from the correct ones extracted by an event-based ensemble.

The goal of this work is threefold. First, we aim to explore whether EP obeys the standard FSS in all dimensions and to explain anomalous FSS through the relationship between the event-based and conventional ensembles. Second, given the simplicity and power of the event-based method, we will use it to determine whether EPs governed by the product and sum rules belong to the same universality class, addressing a long-standing problem in the study of EPs. Third, unlike conventional percolation, where the mean distance between the dynamic pseudocritical point and the infinite-lattice critical point, as well as its fluctuation, vanish at the same rate with increasing system size, we aim to understand the larger fluctuation of the dynamic pseudocritical point in EPs.

Through extensive simulations on hypercubic lattices of side length $L$ across dimensions $d=2$ to $6$ and considering both product and sum rules, we confirm that EPs for both rules exhibit a continuous phase in any dimension. As for the complete graph, the mean of the dynamic pseudocritical point $\langle\mathcal{T}_L\rangle$ deviates from the infinite-lattice criticality by a distance of order $L^{-1/\nu}\sim V^{-1/\nu^*}$, where $V=L^d$ is the system volume and $\nu=\nu^*/d$ is the correlation-length exponent. The distribution of the dynamic pseudo-critical points has a standard deviation scaling as $V^{-\theta}$, with $\theta<1/\nu^*$. Our high-precision estimate of $\theta$ suggests it may take the exact value $\theta=1/2$. We provide a theoretical argument that the dynamic pseudo-critical point asymptotically adheres to the central limit theorem, conforming to a normal (Gaussian) distribution. By fitting our data to the standard FSS ansatz, we accurately determine the percolation threshold and various critical exponents for EPs across dimensions $d=2$ to $6$, and $d=\infty$ (Table~\ref{t1}). Our results underscore the distinct critical exponents observed for EPs under different bond-insertion rules, highlighting their nuanced universalities.

\begin{table}
\caption{The fit results of critical points and critical exponents for EPs on hypercubic lattices of various dimensions. Here, the critical point $T_c$ is defined as the number of inserted bonds at criticality divided by system volume $V=L^d$. To compare systems of different dimensions, including infinite-dimensional systems (complete graphs), the FSS exponents shown here are defined with respect to system volume $V$. As a distinction from the conventional definitions with respect to side length $L$, we use a superscript $*$ for these FSS exponents, so that we have fractal dimension $d_f^*=d_f/d$, and correlation-length exponent $\nu^*=\nu/d$. The exponent $\theta$, describing the fluctuation of the dynamic pseudocritical point, is also defined with respect to system volume $V$. It can be observed that, compared to sum rule, product rule yields a larger critical point $T_c$ (except for $d=2$), a larger correlation-length exponent $\nu^*$, and a smaller fractal dimension $d_f^*$ across all dimensions. For both product and sum rules, the exponent $\theta$ is consistent well with the theoretical value $\theta=1/2$ given by our argument (except for the $d=2$ EP with the product rule, which will be further discussed in Sec.~\ref{sec-ftc}). For comparison, we also list the known percolation thresholds and critical exponents for standard bond percolation~\cite{Paul2001,Wang2013a,Xu2014,Mertens2018,Zhang2021}, for which it has $\theta=1/\nu^*$. With the hyperscaling relation $\tau=1+d/d_f=1+1/d_f^*$, the Fisher exponent $\tau$ is also obtained from the fit results of $d_f^*$. It is demonstrated that EPs for both product and sum rules have the maximum $\tau$ in dimension $d=4$, while $\tau$ monotonously increases with dimensions for standard bond percolation.}
\label{t1}
\begin{ruledtabular}
\begin{tabular}{cclllll}
Model &  $d$ & \multicolumn{1}{c}{$T_c$}   &  \multicolumn{1}{c}{$d_f^*$}  &  \multicolumn{1}{c}{$1/\nu^*$}  &  \multicolumn{1}{c}{$\theta$}  & \multicolumn{1}{c}{$\tau$} \\
\hline
\multirow{5}{1.3cm}{\centering EP \\ Product rule}
& $2$ & $1.053\,126\,4(8)$       & $0.9782(1)$   & $0.5129(1)$  & $0.485(1)$  & $2.022$ \\
& $3$ & $0.966\,300\,6(2)$       & $0.9295(5)$   & $0.616(2)$   & $0.500(1)$  & $2.076$ \\
& $4$ & $0.936\,642\,1(2)$       & $0.911(3)$    & $0.65(2)$    & $0.500(1)$  & $2.098$ \\
& $5$ & $0.923\,282\,8(4)$       & $0.921(2)$    & $0.707(3)$   & $0.500(1)$  & $2.086$ \\
& $6$ & $0.915\,853\,5(1)$       & $0.930(2)$    & $0.724(1)$   & $0.500(1)$  & $2.075$ \\
& $\infty$ & $0.888\,449\,1(2)$  & $0.9349(1)$   & $0.740(2)$   & $0.503(3)$  & $2.070$ \\
\hline
\multirow{5}{1.3cm}{\centering EP \\ Sum rule}
& $2$ & $1.053\,920\,0(6)$       & $0.9797(1)$   & $0.5247(5)$   & $0.501(2)$  &  $2.021$ \\
& $3$ & $0.952\,576\,5(8)$       & $0.9471(1)$   & $0.694(3)$    & $0.500(1)$  &  $2.056$ \\
& $4$ & $0.917\,277\,5(4)$       & $0.945(4)$    & $0.800(6)$    & $0.499(1)$  &  $2.058$ \\
& $5$ & $0.901\,275\,0(5)$       & $0.951(3)$    & $0.83(3)$     & $0.500(1)$  &  $2.052$ \\
& $6$ & $0.892\,386\,4(3)$       & $0.956(3)$    & $0.85(2)$     & $0.501(1)$  &  $2.046$ \\
& $\infty$ & $0.860\,207\,4(1)$  & $0.963(5)$    & $0.85(2)$     & $0.500(1)$  &  $2.038$ \\
\hline
\multirow{5}{1.3cm}{\centering Bond \\ percolation}
& $2$ & $1$          & $91/96$   & $3/8$     & $3/8$    &  $187/91$  \\
& $3$ & $0.746435$   & $0.841$   & $0.571$   & $0.571$  &  $2.189$   \\
& $4$ & $0.640525$   & $0.761$   & $0.365$   & $0.365$  &  $2.313$   \\
& $5$ & $0.590857$   & $0.705$   & $0.349$   & $0.349$  &  $2.417$   \\
& $6$ & $0.565210$   & $2/3$     & $1/3$     & $1/3$    &  $5/2$     \\
& $\infty$ & $1/2$   & $2/3$     & $1/3$     & $1/3$    &  $5/2$
\end{tabular}
\end{ruledtabular}
\end{table}

The remainder of this paper is organized as follows. In Sec.~\ref{sec-model}, we give a brief description of the EP model, event-based ensemble, and observables. Section~\ref{sec-pcp} presents the FSS behaviors of the dynamic pseudocritical points in EPs. Other results for the FSS of EPs in the event-based ensemble are given for dimensions ranging from $2$ to $6$ in Sec.~\ref{sec-scl}. Finally, conclusions are summarized in Sec.~\ref{sec-con}.

\section{Model and method} \label{sec-model}

\subsection{Achlioptas process}

We study EP on a $d$-dimensional hypercubic lattice with side length $L$, resulting in a total system volume of $V\equiv L^d$. Initially ($t=0$), the lattice is empty, with no bonds present. At each time step $t$, a specific bond-insertion rule, to be described later, selects one empty bond to add to the system. As bonds are inserted, the bond density $T\equiv t/V$ increases, eventually leading to the emergence of a giant cluster at the percolation threshold.

A fundamental bond-insertion rule to realize the Achlioptas process is product rule~\cite{Achlioptas2009}. In this rule, two candidate bonds are randomly selected from all empty bonds at each time step. Subsequently, the bond with the smallest product of the sizes of the two clusters containing its two ends is inserted, while the other bond is discarded. Alternatively, instead of the product, the sum of the sizes of the two clusters associated with each candidate bond can be calculated, referred to as sum rule.

\subsection{Event-based ensemble} \label{sec-ebe}

In the Monte Carlo study of FSS, the first step typically involves determining the critical point $T_c$. Subsequently, observables are sampled at or near this fixed bond density $T_c$ across numerous realizations, constituting what we refer to as conventional ensemble. In contrast, the event-based ensemble samples all observables at or around a dynamically determined pseudocritical point $\mathcal{T}_L$, identified by a particular event within a single realization of the percolation process.

Specifically, during an Achlioptas process, we record the sequence of inserted bonds and monitor the one-step incremental size $\Delta(t)$ of the largest cluster, defined as
\begin{equation}
\Delta(t)=\mathcal{C}_1(t+1)-\mathcal{C}_1(t),   \label{eq-delta}
\end{equation}
where $\mathcal{C}_1(t)$ is the size of the largest cluster at time step $t$. As time step $t$ progresses, $\Delta(t)$ generally increases during the subcritical phase, peaks at a certain time step, denoted as $t_{\text{max}}$, and subsequently decreases as the system transitions into the supercritical phase, characterized by the emergence of a large and dense cluster. Consequently, the bond density $\mathcal{T}_{L}$, calculated as $t_{\text{max}}/V$, serves as the pseudocritical point. Given that $\mathcal{T}_{L}$ varies across realizations, we refer to it as the dynamic pseudocritical point. Furthermore, the mean of dynamic pseudocritical points $T_L\equiv\langle\mathcal{T}_L\rangle$ corresponds to a pseudocritical point of traditional definition, maintaining a fixed value for a given system size.

After identifying a dynamic pseudocritical point $\mathcal{T}_{L}$, we reconstruct the percolation configuration at $\mathcal{T}_{L}$ using the recorded bond sequence. Subsequently, we sample physical observables of interest from this critical configuration. The observables in the event-based ensemble are then derived as the means of samples across independent realizations, with $\mathcal{T}_L$ fluctuating in each realization.

It is worth noting that as in Eq.~(\ref{eq-delta}), the event-based ensemble is defined at the dynamic time step $t_{\text{max}}$ before the largest-cluster size $\mathcal{C}_1$ achieves its maximum increment. Alternatively, one can define (equally well) the one-step incremental size of the largest cluster as $\Delta(t)=\mathcal{C}_1(t)-\mathcal{C}_1(t-1)$, for which $t_{\text{max}}$ would then refer to the time step after $\mathcal{C}_1$ achieves its maximum increment. The dynamic pseudocritical points corresponding to these two definitions differ by only one unit, $1/V \sim L^{-d}$, resulting in identical FSS. However, the amplitude of observables could vary significantly when $\mathcal{C}_1$ achieves its maximum increment, leading to different distributions of observables.

\subsection{Observables}

In each realization, the dynamic pseudocritical point $\mathcal{T}_{L}$ is firstly identified, then, the size of the $n$th largest cluster $\mathcal{C}_{n}$ and the number $\mathcal{N}_s$ of clusters of size $s$ are sampled at this dynamic bond density $\mathcal{T}_{L}$. We also identify the bond density $\mathcal{T}_L'$ at which the one-step incremental size $\Delta(t)$ reaches its second maximum, which can also be regarded as another definition of dynamic pseudocritical point. With these samples, we compute the following observables:
\begin{itemize}
  \item The mean dynamic pseudocritical point $T_{L}$, defined as $T_{L} \equiv \langle \mathcal{T}_{L} \rangle$, along with its fluctuation $\sigma(\mathcal{T}_L) \equiv \sqrt{\langle \mathcal{T}_{L}^2 \rangle - \langle \mathcal{T}_{L} \rangle^2}$.
  \item The probability distribution $F(x)$ of $\mathcal{T}_L$ across different realizations.
  \item The size of the largest cluster $C_1$, computed as $C_{1}\equiv\langle\mathcal{C}_{1}\rangle$.
  \item The probability distribution $P(x)$ of $\mathcal{C}_1$ across different realizations.
  \item The susceptibility-like quantity $\chi$, defined as $\chi \equiv \langle \sum_{n\neq1} \mathcal{C}_n^2 \rangle/L^d$, where the sum $\sum_{n}$ encompasses all clusters except the largest one.
  \item The cluster number density $n(s,L)$, defined as $n(s,L)\equiv\langle\mathcal{N}_s\rangle/L^d$, where $\mathcal{N}_s$ denotes the number of clusters of size $s$.
  \item The mean distance $\delta\mathcal{T}$ between $\mathcal{T}_L$ and $\mathcal{T}_L'$, computed as $\delta \mathcal{T}\equiv\langle|\mathcal{T}_L-\mathcal{T}_L'|\rangle$.
\end{itemize}
Here, the angular brackets $\langle \cdot\rangle$ denote averaging across realizations.

\section{Finite-size scaling behaviors of dynamic pseudocritical points}  \label{sec-pcp}

\subsection{The mean dynamic pseudocritical point}

According to the FSS theory, the mean dynamic pseudocritical point $T_L\equiv\langle \mathcal{T}_{L}\rangle$ exhibits an asymptotic scaling behavior described by $T_L-T_c\sim L^{-1/\nu}$, where $T_c$ denotes the infinite-lattice critical point and $\nu$ is the correlation-length exponent. To estimate $T_c$ and $1/\nu$, we fit the data of $T_L$ to the FSS ansatz
\begin{equation}
T_L = T_c+ L^{-1/\nu} (a_0+a_1L^{-\omega_1}+a_2L^{-\omega_2}),  \label{eq-fit1}
\end{equation}
where $\omega_i$ with $i=1$ and $2$ represent for the finite-size corrections. We employ least-squares fits, utilizing a lower cutoff $L_{\text{min}}$ on the data points. Generally, our preferred fit result corresponds to the smallest $L_{\text{min}}$ for which a chi-square per degree of freedom around $1$ is obtained, and increasing $L_{\text{min}}$ should not significantly raise the chi-square per degree of freedom.

When both correction terms in Eq.~(\ref{eq-fit1}) are left unconstrained, we cannot obtain a stable fit. Thus, we set $a_2=0$, thereby considering only one correction term. The stable fit results for the infinite-lattice critical point $T_c$ are presented in Table~\ref{t1}. In comparison to standard bond percolation, it is evident that EP manifests a larger critical point in any dimension for both product and sum rules. This discrepancy suggests that the straightforward rule of EP substantially postpones the onset of the percolation transition in any dimension. While performing these fits, we also observe that $a_0$ changes from a negative value to a positive value when the system reaches or exceeds a dimension of $4$. This implies that $T_L$ eventually approaches $T_c$ from the subcritical side for $d<4$, and from the supercritical side for $d\geq4$.

\begin{table}
\caption{The fit results of critical points for EP in different dimensions. Here, the finite-dimensional system is the hypercubic lattice, and the infinite-dimensional system is the complete graph. In the previous studies, the critical point for hypercubic lattices is usually defined as the number of occupied bonds at criticality normalized by the total number of bonds. Here, we have also converted our estimations of the critical point $T_c$ to this definition. For complete graphs, the critical point is still defined as the number of occupied bonds at criticality normalized by the total number of sites.} \label{t2}
\begin{ruledtabular}
\begin{tabular}{ccll}
$d$  & & \multicolumn{1}{c}{Product rule}         &  \multicolumn{1}{c}{Sum rule}   \\
\hline
\multirow{7}{*}{\centering $2$}
	&  & $0.526\,6(2)$~\cite{Radicchi2010}        &  $0.527\,0(1)$~\cite{Choi2014}      \\
    &  & $0.526\,565(5)$~\cite{Ziff2010}          &  $0.526\,959\,7(6)$~\cite{Wu2022}   \\
    &  & $0.526\,562(3)$~\cite{Grassberger2011}   &   \\
    &  & $0.526\,550(5)$~\cite{Reis2012}          &   \\
    &  & $0.526\,6(1)$~\cite{Choi2014}            &   \\
    &  & $0.526\,563\,6(5)$~\cite{Wu2022}         &   \\
    &\text{Present}    & $0.526\,563\,2(4)$         &  $0.526\,960\,0(3)$ \\
\hline
\multirow{4}{*}{\centering $3$}
    &  & $0.387\,6(2)$~\cite{Radicchi2010}   &  0.3175(1)~\cite{Choi2014}  \\
    &  & $0.322\,096(1)$~\cite{Reis2012}     &   \\
    &  & $0.322\,1(1)$~\cite{Choi2014}	     &   \\
    &\text{Present}    & $0.322\,100\,2(1)$  &  $0.317\,525\,5(4)$  \\
\hline
\multirow{2}{*}{\centering $4$}
    &  & $0.234\,160(3)$~\cite{Reis2012}     &  \\
    &\text{Present}    & $0.234\,160\,5(1)$  &  $0.229\,319\,4(2)$	  \\
\hline
\multirow{2}{*}{\centering $5$}
    &  & $0.184\,656(6)$~\cite{Reis2012}     &  \\
    & \text{Present}    & $0.184\,656\,6(2)$ &   $0.180\,255\,0(2)$	  \\
\hline
\multirow{2}{*}{\centering $6$}
    &  & $0.152\,642(5)$~\cite{Reis2012}     &   \\
    & \text{Present}    & $0.152\,642\,3(1)$ &  $0.148\,731\,1(2)$  \\
\hline
\multirow{6}{*}{\centering $\infty$}
    &      & $0.888\,2(2)$~\cite{Radicchi2010}       &  $0.860\,207(1)$~\cite{Li2023}  \\
    &      & $0.888\,449\,0(5)$~\cite{Lee2011}       &  \\
    &      & $0.888\,449(2)$~\cite{Grassberger2011}  &  \\
    &      & $0.888\,44(2)$~\cite{Fan2012}           &  \\
    &      & $0.888\,449\,4(3)$~\cite{Yang2024}      &  \\
    &  \text{Present}  & 0.888\,449\,1(2)~\cite{Li2023} &  $0.860\,207\,4(1)$ 
\end{tabular}
\end{ruledtabular}
\end{table}

In addition, the critical points $T_c$ of EP in different dimensions are also estimated using other methods. We summarize the reported numerical results in Table~\ref{t2}. In these studies, the critical points for finite dimensions are usually defined as the number of occupied bonds at criticality, normalized by the total number of bonds in the lattice. Thus, we have also converted our fit results for the critical points to this definition. It can be observed that these values are consistent with our fit results within the margins of error, though our estimates have higher precision.

\begin{figure}
\centering
\includegraphics[width=1.0\columnwidth]{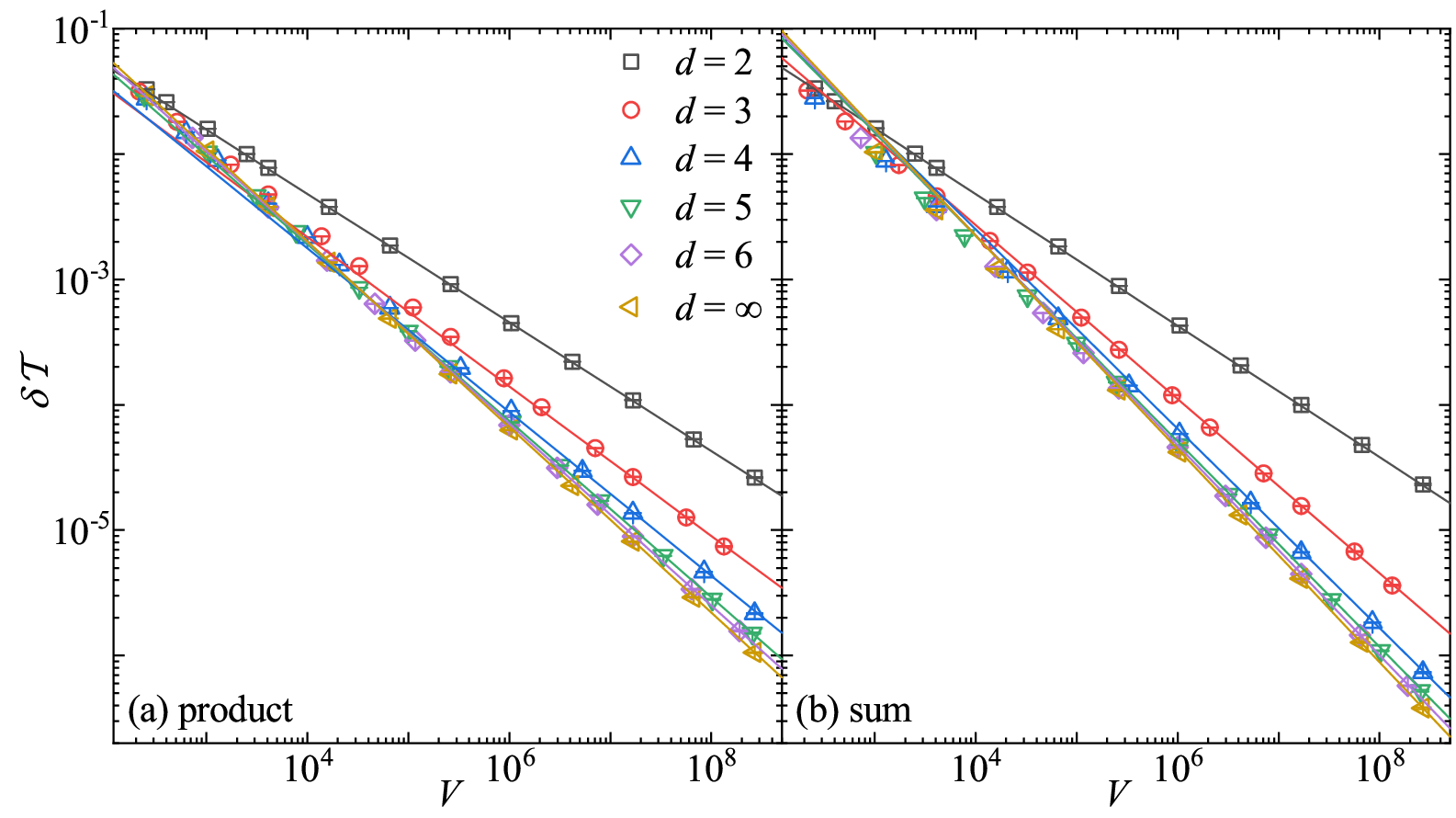}
\caption{(Color online) The mean distance $\delta\mathcal{T}\equiv\langle|\mathcal{T}_L-\mathcal{T}_L'|\rangle$ between the two dynamic pseudocritical points identified by the largest and second-largest one-step incremental size $\Delta(t)$ of the largest cluster, as a function of system volume $V=L^d$. (a) Product rule. (b) Sum rule. Since $\mathcal{T}_L$ and $\mathcal{T}_L'$ follow the same FSS ansatz Eq.~(\ref{eq-fit1}), the mean distance $\delta\mathcal{T}$ between them should exhibit a FSS of $\delta\mathcal{T}\sim L^{-1/\nu}\sim V^{-1/\nu^*}$. Here, the lines represent the scaling $\sim V^{-1/\nu^*}$ with fit results of $1/\nu^*$ in Table~\ref{t1}.} \label{f1}
\end{figure}

By fitting the data of $T_L$ to the scaling ansatz given by Eq.~(\ref{eq-fit1}), we can also determine the correlation-length exponent $1/\nu$. However, an alternative approach can also be employed to ascertain $1/\nu$. This method involves defining another dynamic pseudocritical point $\mathcal{T}_L'$ as the bond density at which the one-step incremental size $\Delta(t)$ reaches its second maximum. Since the dynamic pseudocritical point $\mathcal{T}_L'$ also adheres to the FSS ansatz Eq.~(\ref{eq-fit1}), the difference between the two pseudocritical points should exhibit a scaling behavior of $\delta \mathcal{T}=\langle|\mathcal{T}_L-\mathcal{T}_L'|\rangle\sim L^{-1/\nu}$. In Fig.~\ref{f1}, we plot $\delta\mathcal{T}$ as a function of system volume $V$, where a nice power law can be observed for both product and sum rules in any dimensions. Interestingly, we observe that $\delta\mathcal{T}$ displays weaker finite-size corrections, enabling a more precise determination of the exponent $1/\nu$. Thus, the exponent $1/\nu$ can be estimated by fitting the data to the scaling ansatz
\begin{equation}
\delta\mathcal{T}= L^{-1/\nu} (a_0+a_1L^{-\omega_1}+a_2L^{-\omega_2}).   \label{eq-fit2}
\end{equation}
For $d=2$ and $3$, as well as $d=\infty$ under product rule, stable fits can be obtained by excluding all the correction terms ($a_1=a_2=0$), or including only one correction term with a free exponent $\omega_1$ ($a_2=0$), which yield consistent results. For $d>3$ (except for $d=\infty$ under product rule), stable fits cannot be found with free $\omega_1$ ($a_2=0$). The stable fit results are obtained by excluding all the correction terms with a large $L_{\text{min}}$. We also tried fitting for $d>3$ with a series of fixed $\omega_1$ and $\omega_2$, but the fit results are very sensitive to the values of $\omega_1$ and $\omega_2$.

The fit results are summarized in Table~\ref{t1}. To compare systems of different dimensions, including infinite-dimensional systems (complete graphs), the exponent $\nu^*$ shown in Table~\ref{t1} is defined with respect to system volume $V$, giving $\nu=\nu^*/d$. Notably, these values of $1/\nu^*$ are all larger than those of standard percolation, corroborating the markedly sharp nature of EP in any dimension. Furthermore, it can be observed that for standard bond percolation , the exponent $1/\nu^*$ decreases with dimensions (excluding $d=2$), whereas for EP, it increases with dimensions. This suggests a distinct dimensional dependence between EP and standard bond percolation.

\subsection{The fluctuation of dynamic pseudocritical points}   \label{sec-ftc}

\begin{figure}
\centering
\includegraphics[width=1.0\columnwidth]{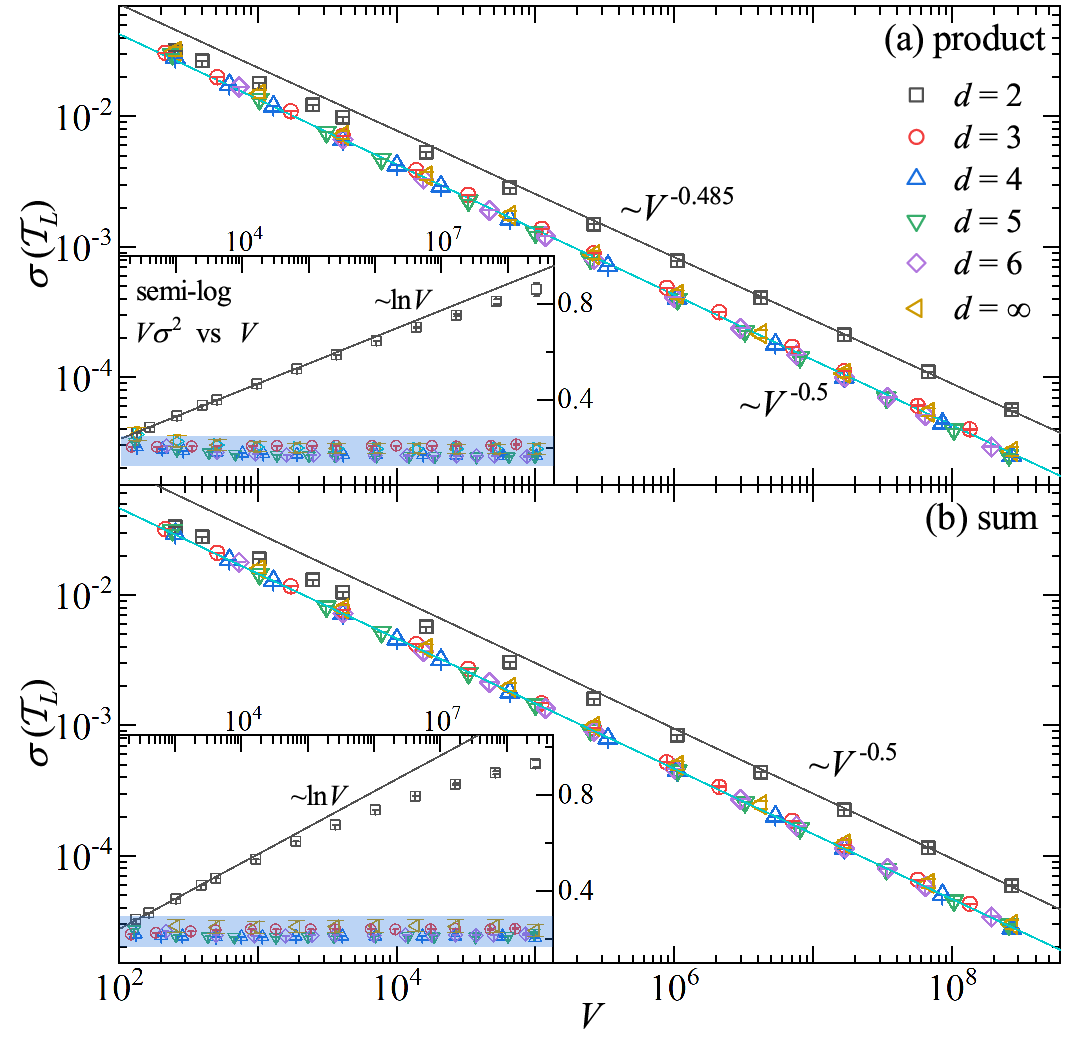}
\caption{(Color online) The fluctuation $\sigma(\mathcal{T}_{L})$ of the dynamic pseudocritical point $\mathcal{T}_{L}$ as a function of system volume $V=L^d$. (a) Product rule. (b) Sum rule. For both product and sum rules, the data of $d\geq3$ nearly collapses on top of a nice scaling $\sigma\sim V^{-1/2}$, indicated by the lines, while the data of $d=2$ show a larger $\sigma(\mathcal{T}_{L})$ with strong finite-size corrections. The fit results suggest the FSS $\sigma(\mathcal{T}_{L})\sim V^{-1/2}$ also holds for sum rule of $d=2$. For product rule, $\sigma(\mathcal{T}_{L})$ for $d=2$ slightly deviates from the scaling $\sigma\sim V^{-1/2}$, and the corresponding line shows the fit result. The insets shows the semi-log plot of the rescaled quantity $V\sigma^2$ as a function of system volume $V$. For $d\geq3$, $V\sigma^2$ rapidly converges to size-independent values indicated by the blue area, suggesting $\sigma\sim V^{-1/2}$. For $d=2$, $V\sigma^2$ also shows a trend toward a size-independent value, but it converges more slowly, as indicated by a negative deviation from logarithmic growth. Thus, the numeric results suggest $\sigma\sim V^{-1/2}$ is a universal property across dimensions and bond-inserting rules.} \label{f2}
\end{figure}

To characterize the FSS behavior of the dynamic pseudocritical point $\mathcal{T}_L$, we further consider its fluctuation, denoted by $\sigma(\mathcal{T}_L)\equiv\sqrt{\langle\mathcal{T}_{L}^2\rangle-\langle\mathcal{T}_{L}\rangle^2}$. In Fig.~\ref{f2}, we plot $\sigma(\mathcal{T}_L)$ as a function of system volume $V=L^d$ for both product and sum rules in dimensions from $d=2$ to $6$, and $d=\infty$. We can find for all the scenarios, $\sigma(\mathcal{T}_L)$ has a FSS of form
\begin{equation}
\sigma(\mathcal{T}_L) \sim V^{-\theta}. \label{eq-plsd}
\end{equation}
To determine the value of the exponent $\theta$, we fit the data of $\sigma$ to the FSS ansatz
\begin{equation}
\sigma(\mathcal{T}_L) = V^{-\theta} (a_0+a_1V^{-\omega_1}+a_2V^{-\omega_2}).  \label{eq-fit3}
\end{equation}
By fixing $a_2=0$ to stabilize the fit, we obtain high-precision results, listed in Table~\ref{t1}.

The fit results of $\theta$ suggest that $\theta<1/\nu^*$ in all finite dimensions for both product and sum rules, consistent with observations in infinite dimensions~\cite{Grassberger2011,Fan2020,Li2023}. Moreover, except EP of product rule in $d=2$, the high-precision values of $\theta$ in Table~\ref{t1} support a universal value $\theta=1/2$, implying that the fluctuation of dynamic pseudocritical points exhibits a universal FSS behavior $\sigma\sim V^{-1/2}$, independent of spatial dimension and bond-insertion rules. This behavior is well illustrated in the insets of Fig.~\ref{f2}, where the rescaled quantity $V\sigma^2$ for $d\geq3$ rapidly approaches a constant as $V$ increases. For $d=2$, the quantity $V\sigma^2$ deviates downward from a straight line ($\sim\ln V$) in the semi-log plot for large systems, suggesting that as $V$ increases, $V\sigma^2$ has the trend to be a $V$-independent value. This trend suggests that $\sigma\sim V^{-1/2}$ may indeed be a universal property for the dynamic pseudocritical point of EPs. The small deviation $1/2-\theta\approx 0.015$ for EP under product rule in $d=2$ could be attributed to finite-size effects not adequately captured in the fit.

\subsection{The normal distribution of dynamic pseudocritical points}

\begin{figure}
\centering
\includegraphics[width=\columnwidth]{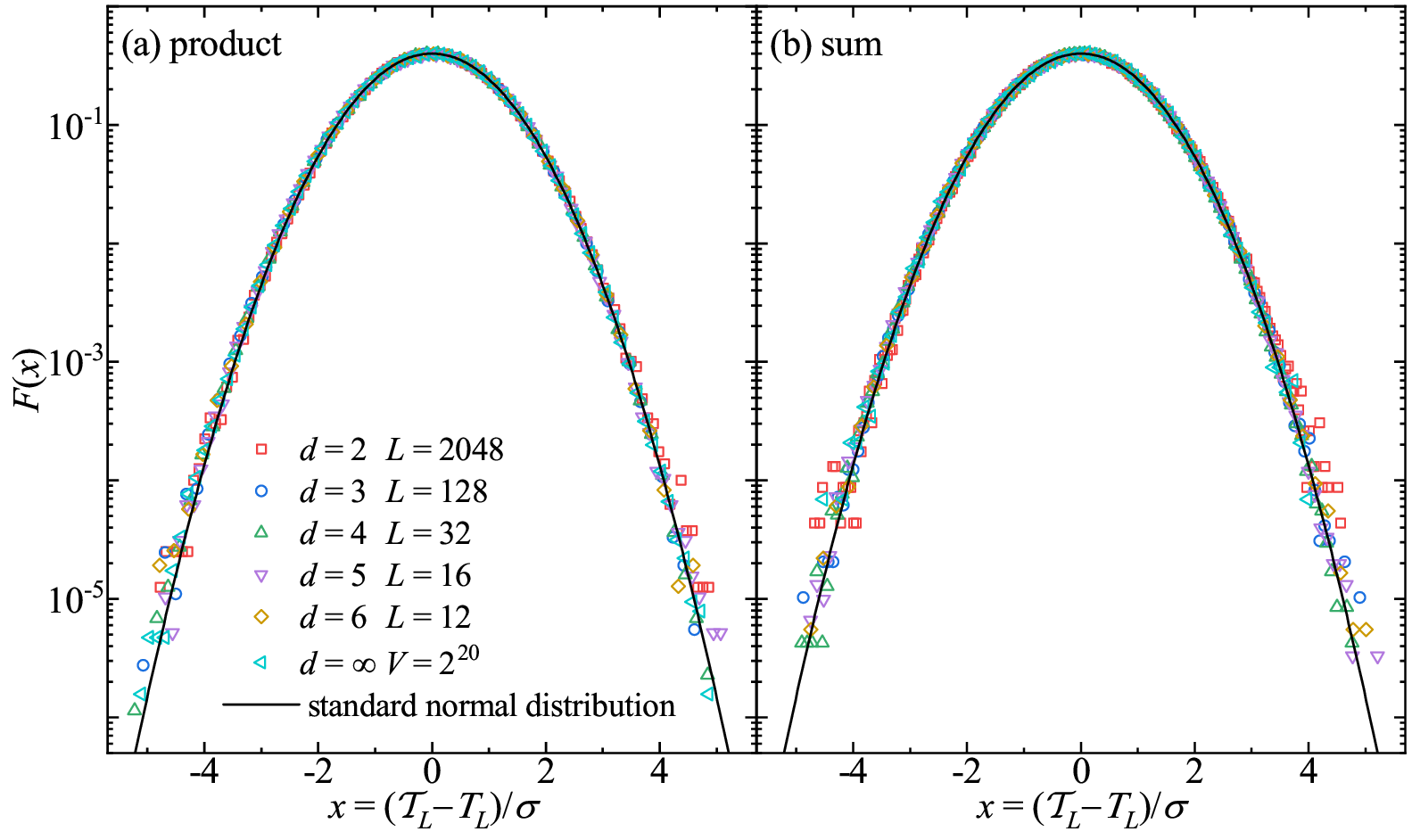}
\caption{(Color online) The probability distribution $F(x)$ of the dynamic pseudocritical point $\mathcal{T}_L$ in dimensions from $d=2$ to $6$, and $d=\infty$. In the simulation, the side length $L$ of hypercubic lattices is chosen to yield a system volume of approximately $V\sim 10^6$. By defining $x\equiv (\mathcal{T}_L-T_L)/\sigma$, where $T_L$ and $\sigma$ represent the mean and standard deviation of $\mathcal{T}_L$, respectively, the simulation results align well with the standard normal distribution, as indicated by the line. The larger fluctuations in the data points of large and small $x$ are due to sparse sampling in these regions.} \label{f3}
\end{figure}

To understand the universal exponent $\theta=1/2$ suggested by Fig.~\ref{f2}, we delve deeper into the distribution of dynamic pseudocritical points. In Fig.~\ref{f3}, the probability distribution $F(x)$ of dynamic pseudocritical points $\mathcal{T}_L$ is plotted for different dimensions, where the logarithmic scale is used to enhance the visibility of the data points that deviate significantly from the mean. Defining $x\equiv (\mathcal{T}_L-T_L)/\sigma$, we observe a remarkable collapse of the data points onto the standard normal distribution, even for very small and very large values of $x$, regardless of spatial dimensions, system sizes, or bond-insertion rules. Although sparse sampling in the regions of extreme $x$ leads to larger fluctuations in the data points, the overall consistency of this collapse provides robust evidence that the dynamic pseudocritical point indeed follows a normal distribution. A conjecture for the normal distribution has also been proposed for EP in infinite dimensions~\cite{Li2023,Feshanjerdi2024}.

To gain insight into these properties of $\mathcal{T}_L$ in EP, we proceed with a theoretical argument. In EP, bonds are selected from a pool of potential candidates by minimizing the product or sum of cluster sizes. Thus, even for high bond densities, the inserted bonds can hardly contribute to the growth of large clusters; rather, they primarily increase the bond density. Therefore, we argue that for the majority of inserted bonds, their contributions to the formation of the giant cluster are unimportant, making them effectively uncorrelated with one another. Consequently, the total number $t_{\text{max}}\sim V$ of inserted bonds at criticality, which is predominantly contributed by these uncorrelated bonds, should asymptotically follow the central-limit theorem, behaving as an extensive quantity with a fluctuation of $\sigma(t_{\text{max}})\sim\sqrt{V}$. As a result, the distribution of dynamic pseudocritical points, defined as $\mathcal{T}_L\equiv t_{\text{max}}/V$, should conform to a normal distribution, as suggested by the central-limit theorem. Moreover, the fluctuation of dynamic pseudocritical points can be readily understood as $\sigma(\mathcal{T}_L)=\sigma(t_{\text{max}})/V\sim V^{-1/2}$, explaining the consistent values of $\theta=1/2$ listed in Table~\ref{t1}.

It is important to emphasize that this argument is specific to the bond-insertion rule of EP. Without a bond-insertion rule that suppresses the growth of large clusters as EP, the inserted bonds will significantly influence the formation of a giant cluster as bond density increases, leading to correlations among the inserted bonds. These correlations deviate from the assumptions underpinning the central-limit theorem. Therefore, the argument presented here for EP does not extend to other percolation models. The diverse numerical results reported for various percolation models in Ref.~\cite{Feshanjerdi2024} also indicate that the dynamic pseudocritical point does not generally follow a normal distribution.

\section{Finite-size scaling of explosive percolation}  \label{sec-scl}

\subsection{Difference between the conventional ensemble and the event-based ensemble}

\begin{figure}
\centering
\includegraphics[width=1.0\columnwidth]{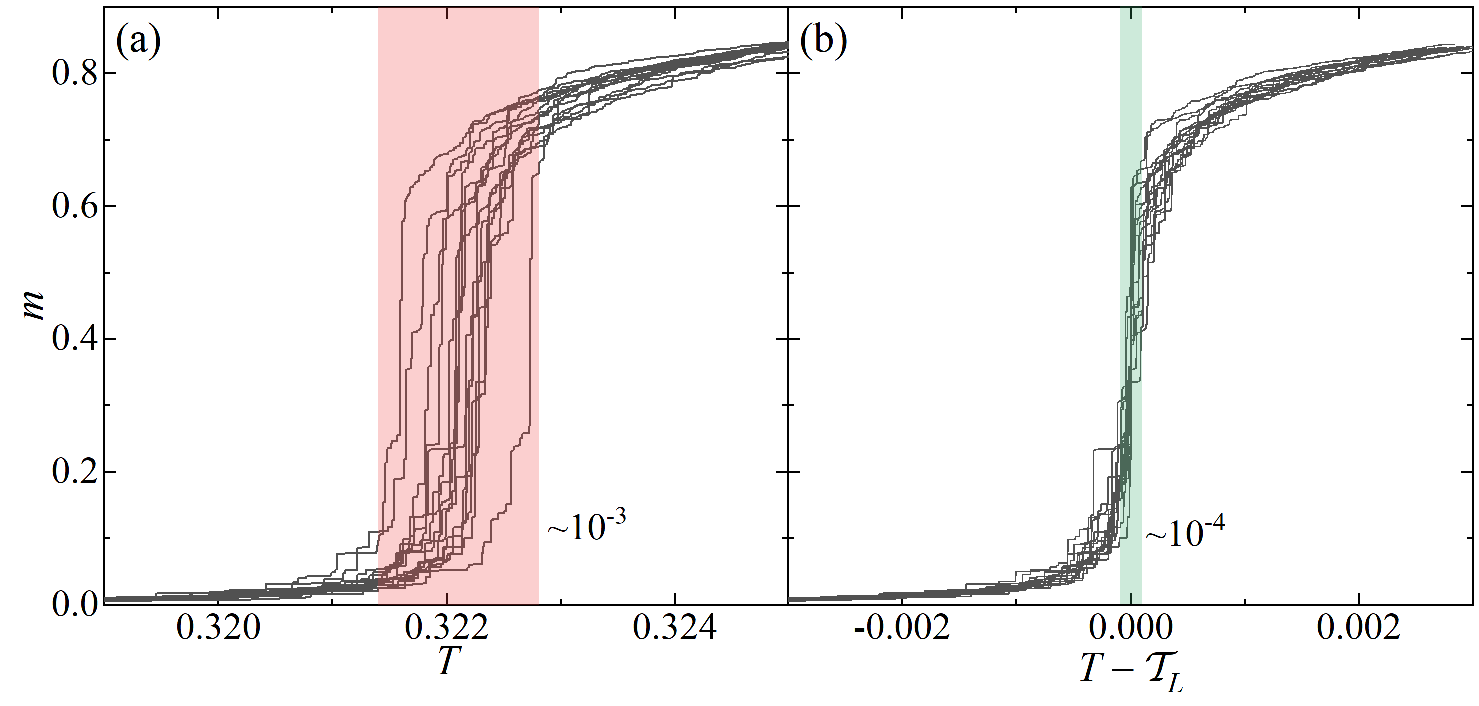}
\caption{(Color online) Illustration of the substantial sample-to-sample fluctuation of EP. The simulations are conducted on a cubic lattice with a side length of $L=64$. (a) The order parameter $m\equiv\mathcal{C}_1/V$ as a function of bond density $T$ for $15$ independent realizations. The red area highlights the sample-to-sample fluctuation of approximately $\mathcal{O}(10^{-3})$, corresponding to $\mathcal{O}(V^{-\theta})$ with $V=64^3$ and $\theta=1/2$. (b) The same $15$ realizations as in (a) after aligning their $\mathcal{T}_{L}$. The green area highlights the scaling window $L^{-1/\nu}\sim V^{-1/\nu^*}\sim 10^{-4}$ around $\mathcal{T}_{L}$ from $V=64^3$ and $1/\nu^*=0.616$. These figures underscore that the sample-to-sample fluctuation $\mathcal{O}(V^{-\theta})$ surpasses the scaling window $\mathcal{O}(L^{-1/\nu})$, emphasizing the necessity to derive accurate critical behaviors at $\mathcal{T}_{L}$ rather than $T_c$.} \label{f4}
\end{figure}

As illustrated in Table~\ref{t1}, the exponent $\theta$ consistently falls below $1/\nu^*$, indicating that in a single realization, the critical point $T_c$ could deviate significantly from the dynamic pseudocritical point $\mathcal{T}_L$. To visually illustrate this phenomenon, we take EP on the cubic lattice of $L=64$ as an example and plot the order parameter $m\equiv\mathcal{C}_1/V$ of independent realizations against bond density $T$ in Fig.~\ref{f4} (a). It is evident that the system fluctuates within a considerable range of order $V^{-\theta}\sim 10^{-3}$ across different realizations, as indicated by red in Fig.~\ref{f4} (a). On the other hand, these curves exhibit remarkable similarity, differing only in their displacement. When they are shifted using a realization-dependent factor $\mathcal{T}_{L}$, they almost entirely overlap, where the fluctuation is smaller and scales as $L^{-1/\nu}\sim 10^{-4}$, as indicated by green in Fig.~\ref{f4} (b). This intriguing property persists across dimensions, suggesting that the relative fluctuation among different EP realizations remains consistent and is solely dependent on the scaling window $\mathcal{O}(L^{-1/\nu})$.

\begin{figure}
\centering
\includegraphics[width=1.0\columnwidth]{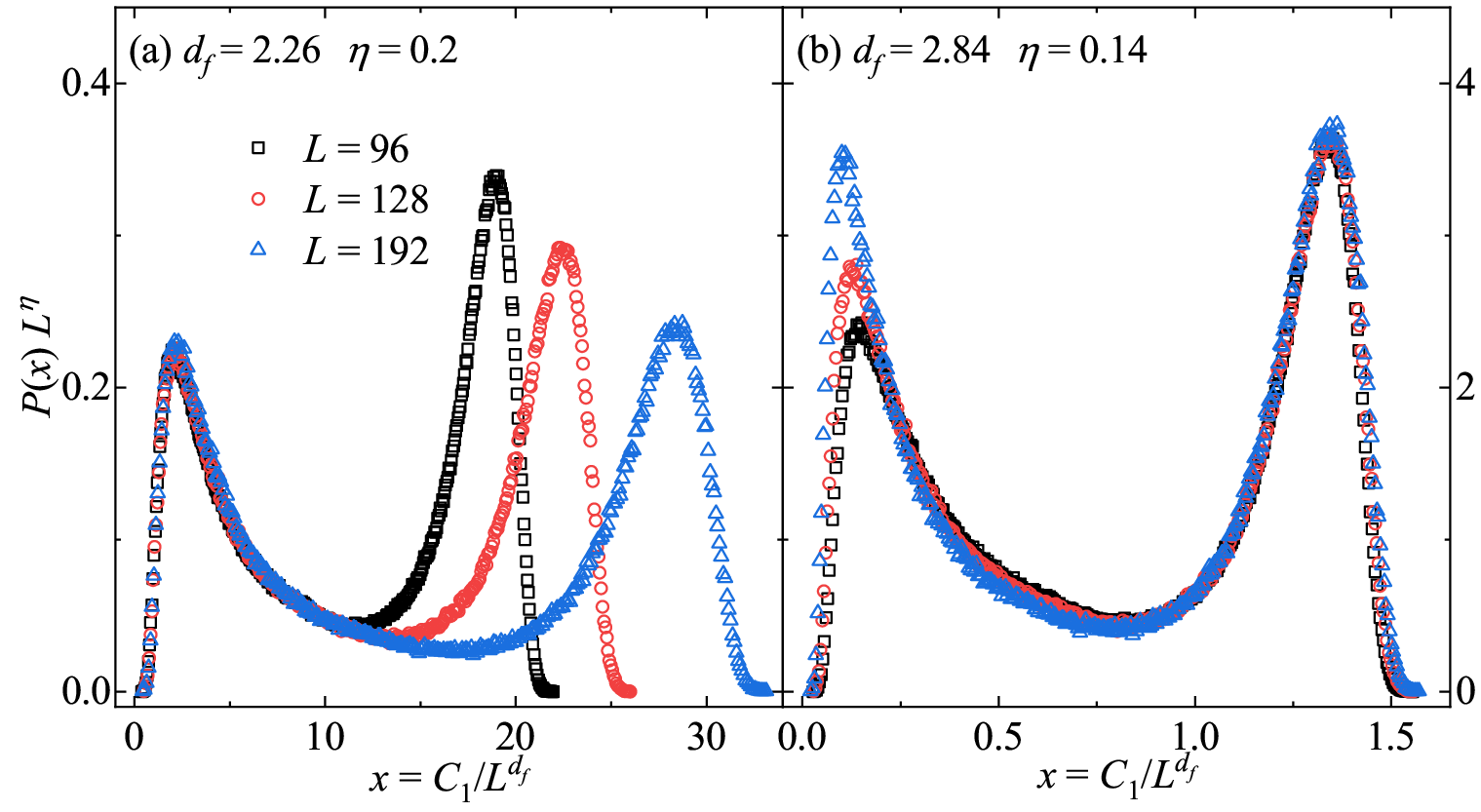}
\caption{(Color online) The distribution $P(x)$ of the largest-cluster size $C_1$ at $T_c=0.9663006$ for EP under product rule in dimension $d=3$. The standard FSS theory suggests that $P(x)$ of different system sizes can collapse on top of each other by defining $x\equiv C_1/L^{d_f}$, where $d_f$ is the fractal dimension of the largest-cluster size $C_1$. Here, however, the distribution $P(x)$ is bimodal, and data collapse of $P(x)$ cannot be achieved simultaneously in the whole range of $x$, regardless of the choice of $d_f$ in $x\equiv C_1/L^{d_f}$. To roughly collapse the data of different system sizes near the two peaks, distribution $P(x)$ should be rescaled as $P(x)L^\eta$, with different $d_f$ and $\eta$ for the two peaks.} \label{f5}
\end{figure}

Consequently, it is more appropriate to extract FSS behaviors at the dynamic pseudocritical point $\mathcal{T}_L$~\cite{Li2023,Yang2024}, which refers to event-based ensemble. In contrast, at any fixed bond density near criticality, i.e., in the conventional ensemble, independent realizations of EP may yield results in the supercritical, critical, or subcritical phases, exhibiting distinct behaviors in observables. Blindly averaging these data will introduce a mixture effect, thereby complicating the application of traditional FSS methods. For instance, as $T_c$ can lie either at the supercritical or subcritical sides of $\mathcal{T}_L$, where the largest clusters differ significantly in size, the largest-cluster size $C_1$ sampled at $T_c$ generally presents a bimodal distribution~\cite{Grassberger2011,Tian2012,Li2023}.

In Fig.~\ref{f5}, we provide an example of the bimodal distribution $P(x)$ of the largest-cluster size $C_1$ at $T_c$ for $d=3$. According to standard FSS theory, $P(x)$ for different system sizes should collapse onto a single curve by defining $x\equiv C_1/L^{d_f}$, where $d_f$ is the fractal dimension of the largest-cluster size $C_1$. However, due to the mixture effect at $T_c$, this data collapse cannot be achieved across the entire range of $x$, regardless of the choice of $d_f$ in $x\equiv C_1/L^{d_f}$, indicating the limitations of the conventional ensemble in capturing a well-defined critical behavior for EP. Similar to EP in infinite dimensions~\cite{Li2023}, the multiple fractal dimensions in Fig.~\ref{f5} can also be related to the clean FSS extracted at $\mathcal{T}_L$ using the crossover FSS theory~\cite{Li2024}. The details will be provided after presenting the clean FSS in the next section.

\subsection{Finite-size scaling in the event-based ensemble}

\subsubsection{Fractal dimension}

\begin{figure}
\centering
\includegraphics[width=1.0\columnwidth]{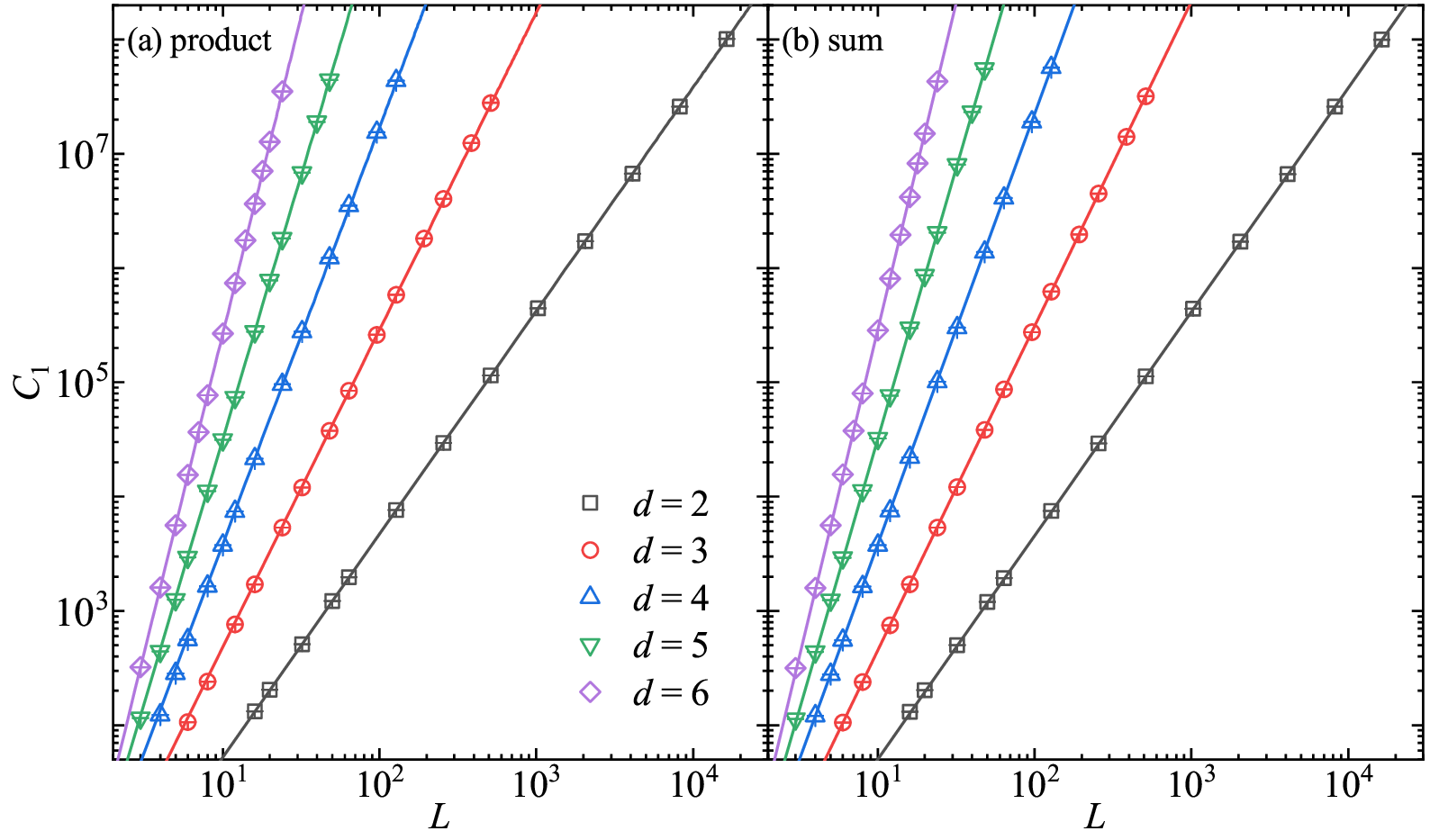}
\caption{(Color online) The size $C_1$ of the largest cluster as a function of side lengths $L$ for hypercubic lattices across dimensions $2$ to $6$. The lines overlaid on the data points represent the scaling $C_1\sim L^{d_f}$ with fit results of $d_f^*=d_f/d$ in Table~\ref{t1}. (a) Product rule. (b) Sum rule. } \label{f6}
\end{figure}

As an illustration of the clean FSS in the event-based ensemble, we present the clear power-law growth of the largest-cluster size $C_1$ with increasing side length $L$, depicted as $C_1\sim L^{d_f}$ in Fig.~\ref{f6}, where $d_f$ is the fractal dimension. To determine the fractal dimension, we fit the $C_1$ data to the FSS ansatz
\begin{equation}
C_1 = L^{d_f} (a_0+a_1L^{-\omega_1}+a_2L^{-\omega_2}),  \label{eq-fit4}
\end{equation}
where the terms of $L^{-\omega_i}$ with $i=1,2$ are the finite-size corrections. For dimensions $d=2$ and $3$, as well as $d=\infty$ under product rule, a stable fit can be achieved by considering only one correction term in Eq.~(\ref{eq-fit4}), with the correction exponent $\omega_1$ left free. However, for dimensions $d>3$ (except for $d=\infty$ under product rule), obtaining a stable fit with free correction exponents $\omega_1$ and $\omega_2$ proves challenging. Consequently, we perform multiple fits with different fixed values of $\omega_1$ and $\omega_2$. The final estimation of the fractal dimension accounts for these varied fits, resulting in a lower precision compared to dimensions $d=2$ and $3$. Notably, excluding the two correction terms from Eq.~(\ref{eq-fit4}), consistent fractal dimensions are also observed with a larger $L_{\text{min}}$.

The stable fit results for fractal dimensions are summarized in Table~\ref{t1}, where the volume fractal dimension is defined as $C_1\sim V^{d_f^*}$ with $d_f^*=d_f/d$. It can be observed that a larger fractal dimension than standard bond percolation in any finite dimension for both product and sum rules, so that, it underscores the explosive yet continuous nature of EP. Moreover, the fractal dimension $d_f^*$ of standard bond percolation monotonically decreases with dimensions, while for EP, $d_f^*$ reaches a minimum value for $d=4$, also indicating a special dimension-dependent property for EP.

\begin{figure}
\centering
\includegraphics[width=\columnwidth]{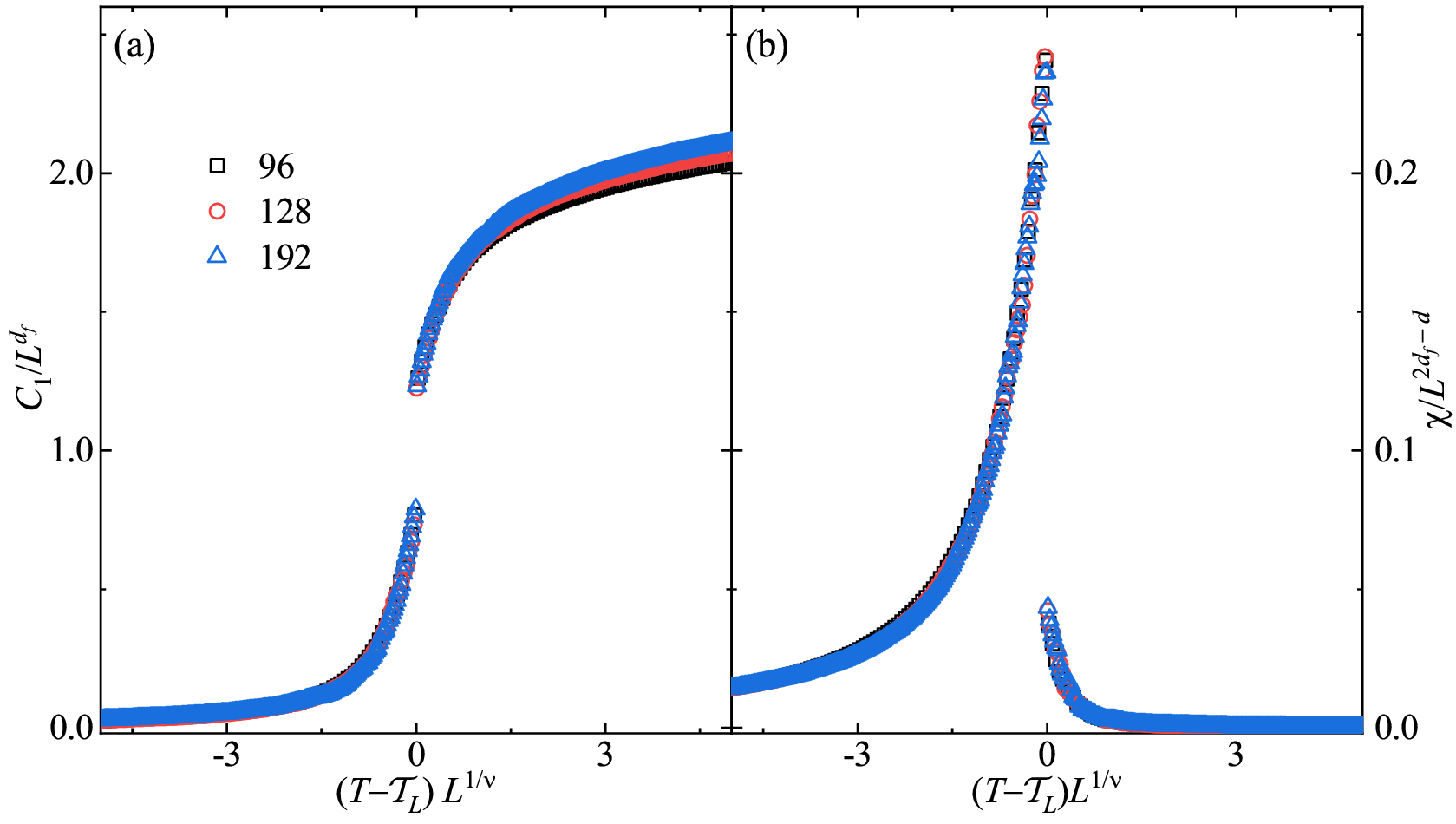}
\caption{(Color online) A demonstration of the FSS behaviors of EP around the dynamic pseudocritical point $\mathcal{T}_L$, with the example of the product rule for dimension $d=3$. (a) The rescaled largest-cluster size $C_1/L^{d_f}$ as a function of $\epsilon L^{1/\nu}=(T-\mathcal{T}_L)L^{1/\nu}$ for different side lengths $L$. (b) The rescaled susceptibility $\chi/L^{2d_f-d}$ as a function of $\epsilon L^{1/\nu}=(T-\mathcal{T}_L)L^{1/\nu}$ for different side lengths $L$. In the plot, the fit results of $d_f^*=d_f/d$ and $1/\nu^*=1/d\nu$ in Table~\ref{t1} are used. The discontinuity at $\epsilon=T-\mathcal{T}_L=0$ arises from the event-based definition of $\mathcal{T}_L$.} \label{f7}
\end{figure}

\begin{figure*}
\centering
\includegraphics[width=2.0\columnwidth]{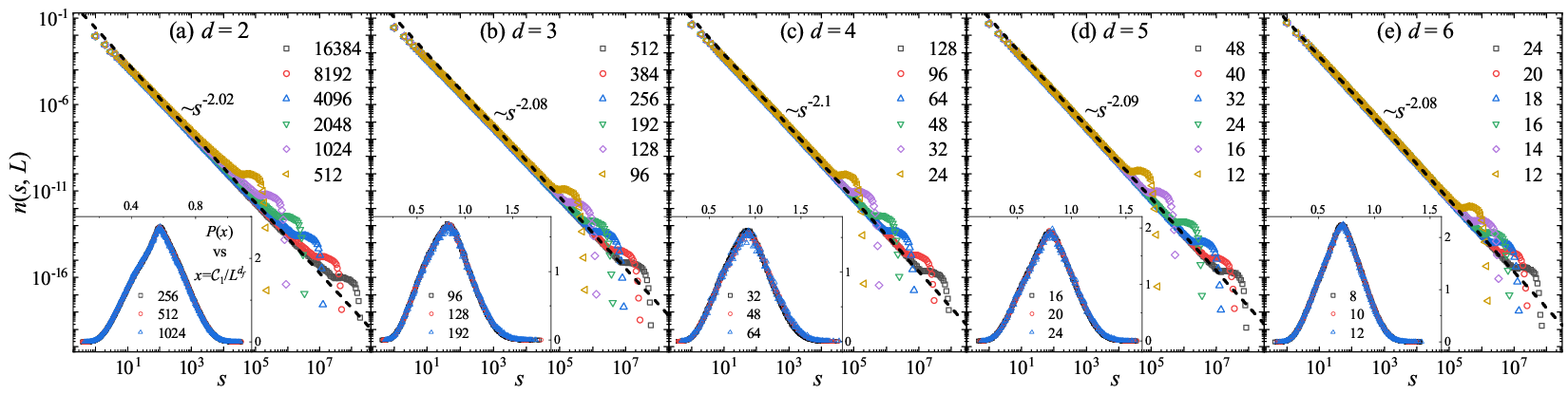}
\caption{(Color online) The cluster number density $n(s,L)$ of EP under product rule is plotted for various side lengths $L$ across dimensions $2$ to $6$. The dashed lines represent the Fisher exponent $\tau=1+d/d_f=1+1/d_f^*$ calculated using the fractal dimensions $d_f^*$ listed in Table~\ref{t1}. Although the data points of small clusters have a slight deviation from the dashed line, especially for $d=2$ and $3$, we note a clear trend towards satisfying the scaling $\sim s^{-\tau}$ as $L\to\infty$, consistent with the fit values of $d_f^*$. Notably, the Fisher exponent shown here reaches its maximum value in $d=4$, while for standard percolation, the Fisher exponent monotonically increases with spatial dimensions and reaches the maximum value of $\tau=5/2$ above the upper critical dimension $d_u=6$. The insets show the probability distribution $P(x)$ of the largest-cluster size $\mathcal{C}_1$ in the event-based ensemble for systems of different dimensions $d$ and side lengths $L$. By defining $x\equiv\mathcal{C}_1/L^{d_f}$ with the fractal dimensions $d_f^*=d_f/d$ given in Table~\ref{t1}, the data of different system sizes collapse onto each other remarkably well.} \label{f8}
\end{figure*}

By applying standard FSS theory around $\mathcal{T}_L$, the behaviors of the largest-cluster size $C_1$ and the susceptibility $\chi$ are represented as $C_1=L^{d_f}\tilde{C_1}(\epsilon L^{1/\nu})$ and $\chi=L^{2d_f-d}\tilde{\chi}(\epsilon L^{1/\nu})$, where $\epsilon=T-\mathcal{T}_L$. To demonstrate these FSS behaviors, we take EP under product rule in dimension $d=3$ as an example, and plot $C_1/L^{d_f}$ and $\chi/L^{2d_f-d}$ as a function of $\epsilon L^{1/\nu}$ for different $L$, as shown in Fig.~\ref{f7}. It can be observed that near criticality, the $C_1$ and $\chi$ data for different system sizes collapse onto a single curve, suggesting that EP obeys standard FSS theory in the event-based ensemble. Different from the conventional ensemble, the observables in Fig.~\ref{f7} are discontinuous at $\epsilon=T-\mathcal{T}_L=0$, arising from the event-based definition of $\mathcal{T}_L$.

To further demonstrate the clean FSS in the event-based ensemble, we illustrate the probability distribution $P(x)$ of the largest-cluster size $\mathcal{C}_1$ for EP under product rule in the insets of Fig.~\ref{f8}. Here, $x$ is defined as $\mathcal{C}_1/L^{d_f}$ using the fractal dimensions $d_f=dd_f^*$ from the fit result $d_f^*$ in Table~\ref{t1}. Remarkably, the data from various system sizes collapse onto each other well, reinforcing the clean FSS of EP in the event-based ensemble, with the high-precision $d_f$. It is worth noting that in the conventional ensemble, the probability distribution $P(x)$ exhibits a bimodal distribution at a fixed bond density, as shown in Fig.~\ref{f5} for $d=3$, and Refs.~\cite{Grassberger2011,Tian2012,Li2023} for infinite-dimensional systems.

In addition, as mentioned in Sec.~\ref{sec-ebe}, the dynamic pseudocritical point can alternatively be defined as the bond density after $\mathcal{C}_1$ achieves its maximum increment, or as the bond density that $\mathcal{C}_1$ achieves its second maximum increment. For these pseudocritical points, the same FSS behavior can be observed, as they all fall within the scaling window $\mathcal{O}(L^{-1/\nu})$. It is worth noting, however, that at these pseudocritical points, the $\mathcal{C}_1$ distribution could be different from those in the insets of Fig.~\ref{f8}. Despite this, a robust data collapse can still be achieved using the same fractal dimensions. Examples for EP in infinite-dimensional systems can be found in Refs.~\cite{Zhu2017,Fan2020,Li2023}.

The crossover FSS theory~\cite{Li2023,Li2024} gives that at bond densities $T^\pm=\mathcal{T}_L\pm aL^{-\lambda}$ ($\lambda<1/\nu$) outside the scaling window, where $a$ is $L$-independent, the fractal dimension of the largest cluster takes the forms
\begin{align}
d_f^+ &= d-(d-d_f)\lambda\nu,  \label{eq-dfp}   \\
d_f^- &= d_f\lambda\nu,        \label{eq-dfm}
\end{align}
where $+$ and $-$ refer to the supercritical and subcritical sides, respectively. As the large fluctuation of $\mathcal{T}_L$, the ensemble defined at $T_c$ should contain samples of multiple fractal dimensions corresponding to a continuous change of $\lambda\in[d\theta,1/\nu]$ in Eqs.~(\ref{eq-dfp}) and (\ref{eq-dfm}). The case $\lambda=1/\nu$ corresponds to the FSS in the scaling window, $d_f^+=d_f^-=d_f$, as listed in Table~\ref{t1}. For the case $\lambda=d\theta=d/2$, taking EP under product rule in $d=3$ as an example, which has $d_f=dd_f^*=2.7885$, and $1/\nu=d/\nu^*=1.848$, Eqs.~(\ref{eq-dfp}) and (\ref{eq-dfm}) give $d_f^-\approx 2.26$ and $d_f^+\approx 2.83$. These two fractal dimensions just explain the ones used to collapse the data near the two peaks in Fig.~\ref{f5}, respectively.

\subsubsection{Cluster number density}

To demonstrate the scaling behaviors of all clusters, we examine the cluster number density $n(s,L)$, defined as the count of clusters of size $s$ normalized by the system volume. In critical finite systems, the cluster number density follows the form $n(s,L)=s^{-\tau}\tilde{n}(s/L^{d_f})$, where $\tilde{n}(s/L^{d_f})$ is a universal function, and $\tau$ denotes the Fisher exponent. As $L\to\infty$, the universal function trends to be a constant, resulting in the cluster number density exhibiting a scaling behavior $n(s,\infty)\sim s^{-\tau}$.

In Fig.~\ref{f8}, we observe the expected power-law distribution of $n(s,L)$ for $L\to\infty$ across all dimensions for EP of product rule, indicating the well-defined nature of the cluster number density in the event-based ensemble. For sum rule, similar phenomena can be also observed. By employing the fit results of fractal dimensions, we can find the Fisher exponent using the hyperscaling relation $\tau=1+d/d_f=1+1/d_f^*$ (Table~\ref{t1}), represented by the dashed lines in Fig.~\ref{f8}. While the simulation results exhibit slight deviations from these dashed lines for small clusters, especially for $d=2$ and $3$, a clear trend, towards satisfying the scalings consistent with the fit values of $d_f^*$, can be observed as $L\to\infty$. This reaffirms the validity of our approach in determining critical exponents through the event-based ensemble.

Notably, for standard percolation, the Fisher exponent $\tau$ typically increases monotonically with spatial dimension and reaches a maximum value of $\tau=5/2$ above the upper critical dimension $d_u=6$~\cite{Stauffer1991}. However, the fit results in Table~\ref{t1} reveal a distinct and universal behavior for EPs: the Fisher exponent reaches a maximum value in dimension $d=4$. Furthermore, compared to standard percolation, EP always exhibits a smaller Fisher exponent, regardless of spatial dimensions and bond-insertion rules. This is because the core mechanism of EP, namely, suppressing further growth of large clusters, results in an increase in the number of large clusters, accompanied by a decrease in the number of small clusters.

\subsection{Different universalities for product and sum rules}

\begin{figure}
\centering
\includegraphics[width=\columnwidth]{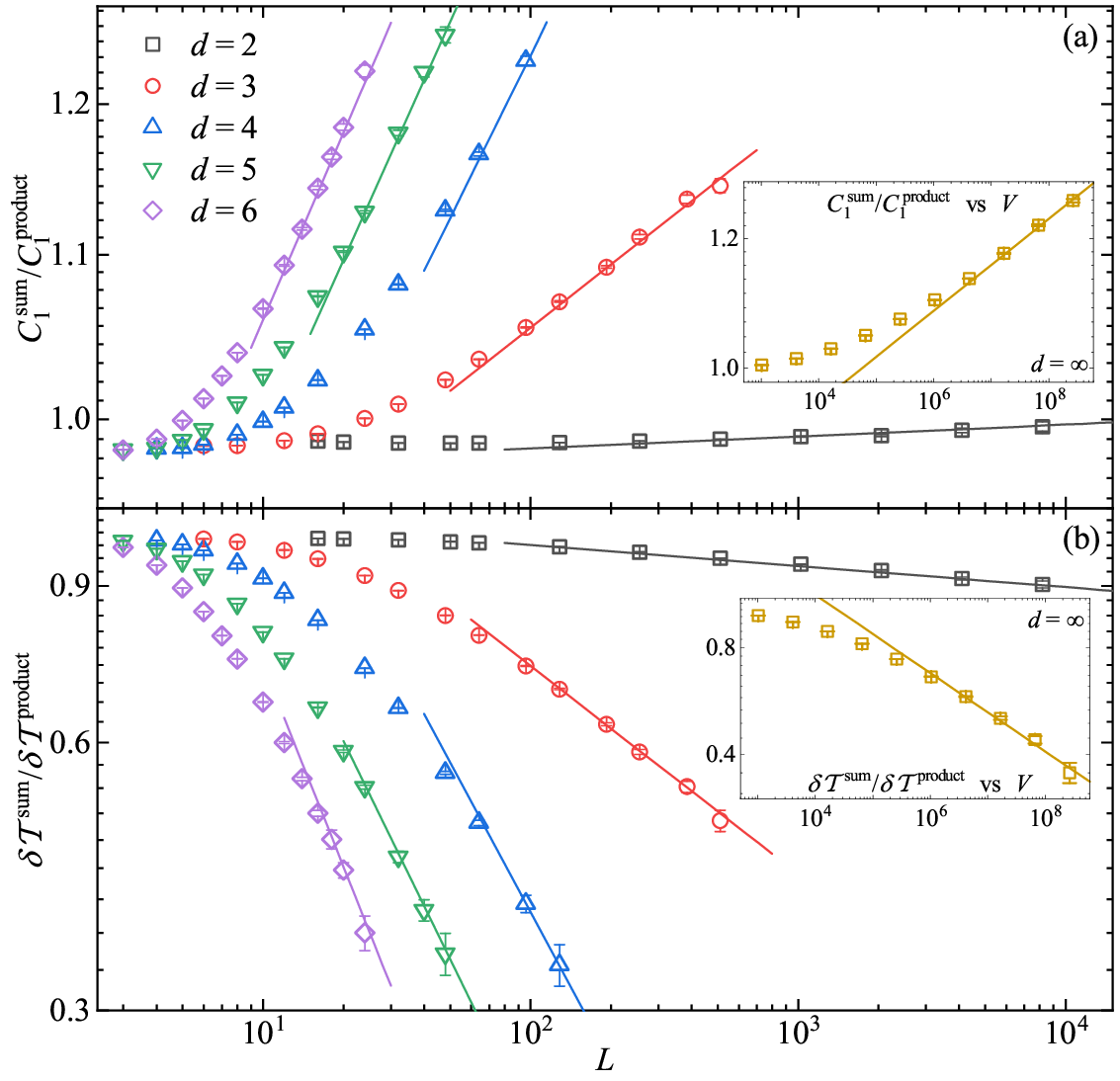}
\caption{(Color online) The FSS to demonstrate different universalies of EPs under product and sum rules.
(a) The ratio $C_1^{\text{sum}}/C_1^{\text{product}}$ as a function of side length $L$. As $L$ increases, a scaling behavior $C_1^{\text{sum}}/C_1^{\text{product}}\sim L^{d_f^{\text{sum}}-d_f^{\text{product}}}$ can be observed, as indicated by the lines with $d_f^*=d_f/d$ given in Table~\ref{t1}. The inset shows the result for $d=\infty$, where the line represents the scaling $C_1^{\text{sum}}/C_1^{\text{product}}\sim V^{d_f^{*\text{sum}}-d_f^{*\text{product}}}$ with $d_f^*$ given in Table~\ref{t1}. (b) The ratio $\delta\mathcal{T}^{\text{sum}}/\delta\mathcal{T}^{\text{product}}$ as a function of side length $L$. As $L$ increases, a scaling behavior $\delta\mathcal{T}^{\text{sum}}/\delta\mathcal{T}^{\text{product}}\sim L^{-1/\nu^{\text{sum}}+1/\nu^{\text{product}}}$ can be observed, as indicated by the lines with $1/\nu^*=1/d\nu$ given in Table~\ref{t1}. The inset shows the result for $d=\infty$, where the line represents the scaling $\delta\mathcal{T}^{\text{sum}}/\delta\mathcal{T}^{\text{product}}\sim V^{-1/\nu^{*\text{sum}}+1/\nu^{*\text{product}}}$ with $1/\nu^*$ given in Table~\ref{t1}.} \label{f9}
\end{figure}

Comparing the critical exponents of EPs under sum and product rules (Table~\ref{t1}) reveals several key properties. Firstly, the exponent $\theta$ for both the two rules aligns with the theoretical prediction $\theta=1/2$ from our argument, underscoring a universal behavior of EP, irrespective of the specific bond-insertion rule employed. Secondly, except in dimension $d=2$, where EPs under the two rules occur almost at the same threshold, the critical point $T_c$ for sum rule is consistently smaller than that of product rule, suggesting product rule has a stronger inhibitory effect on the growth of large clusters. Thirdly, we observe that sum rule generally exhibits a smaller correlation-length exponent $\nu$ and a larger fractal dimension $d_f$ compared to product rule, indicating that EP under sum rule manifests a sharper transition. Fourthly, owing to the high-precision critical exponents for both rules, notable disparities in critical exponents between sum and product rules are observed, surpassing the error bars. This suggests that EPs under these two rules may belong to distinct universality classes, consistent with observations in the infinite-dimensional EP~\cite{Li2023}.

To illustrate the distinguishable universalities of EPs under product and sum rules, we plot the ratio of the largest-cluster size $C_1$ for product and sum rules in Fig.~\ref{f9} (a). A nice power-law growth of $C_1^{\text{sum}}/C_1^{\text{product}}\sim L^{d_f^{\text{sum}}-d_f^{\text{product}}}$ can be observed for large systems, where the exponents of the scalings are consistent with the fit results in Table~\ref{t1}. This suggests that EPs under product and sum rules possesses their own distinct fractal dimensions. Similarly, by plotting $\delta\mathcal{T}^{\text{sum}}/\delta\mathcal{T}^{\text{product}}$ as a function of $L$, a FSS of $\sim L^{-1/\nu^{\text{sum}}+1/\nu^{\text{product}}}$ can be also observed, as shown in Fig.~\ref{f9} (b). For infinite-dimensional systems, similar behaviors can be also observed, see the insets of Fig.~\ref{f9}. Consequently, for any dimensions, EPs under the product and sum rules should belong to different universalities.

\subsection{Upper critical dimension}

\begin{figure}
\centering
\includegraphics[width=\columnwidth]{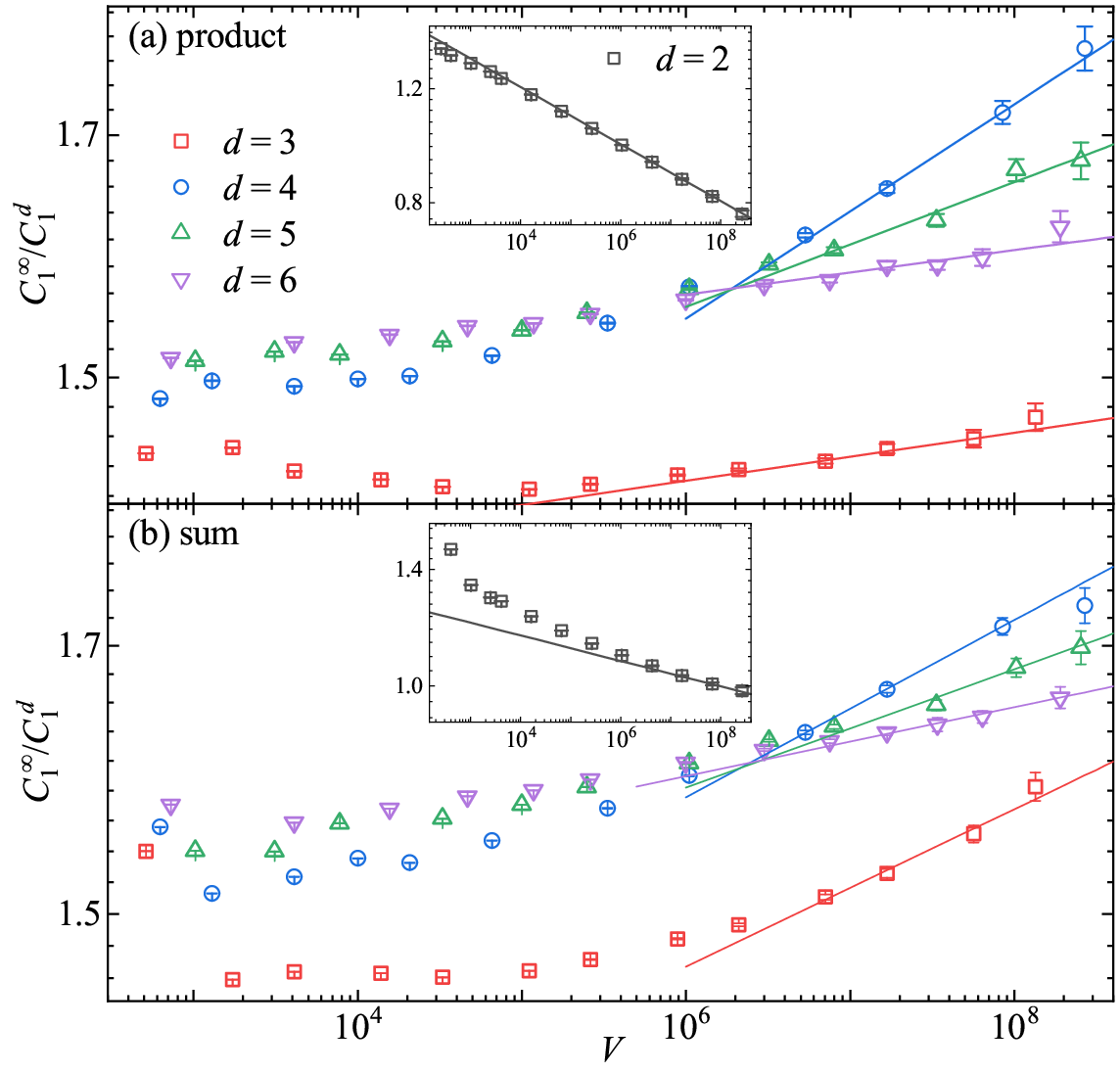}
\caption{(Color online) The FSS analysis showing the change in fractal dimensions with varying dimensions. The ratio $C_1^{\infty}/C_1^{d}$ is plotted as a function of system volume $V$ for product rule (a) and sum rule (b). Here, $C_1^{\infty}$ and $C_1^{d}$ represent the largest-cluster size in infinite and $d$-dimensions, respectively. The line indicates the scaling $\sim V^{d_f^{*\infty}-d_f^{*d}}$, with $d_f^*$ values listed in Table~\ref{t1}. The inset provides a separate plot for $d=2$.} \label{f10}
\end{figure}

From Table~\ref{t1}, we observe that EP exhibits a different dimension-dependence compared to standard bond percolation. The exponent $1/\nu^*$ increases with dimension for both product and sum rules, indicating a sharper phase transition as dimensions increase. In contrast, standard bond percolation shows a smaller $1/\nu^*$ at higher dimensions. Interestingly, the fractal dimension $d_f^*$ of EP reaches a minimum value at $d=4$, whereas for standard percolation, $d_f^*$ decreases with increasing dimensions and attains a minimum value of $d_f^*=2/3$ at the upper critical dimension $d_u=6$.

Within the error margins, EP also demonstrates the same critical exponents for $d=6$ and $d=\infty$ (Table~\ref{t1}). To numerically examine the dimension-dependent property of EP, we present the FSS of the ratio $C_1^{\infty}/C_1^{d}$ in Fig.~\ref{f10}, where $C_1^{\infty}$ and $C_1^{d}$ represent the largest-cluster size in infinite and $d$-dimensions, respectively. We observe consistent properties for EPs under both product and sum rules. The fractal dimension $d_f^*$ reaches its maximum value at $d=2$, causing $C_1^{\infty}/C_1^{d=2}$ to vanish as system size increases (see the insets of Fig.~\ref{f10}). For $d>2$, the ratio $C_1^{\infty}/C_1^{d}$ grows with increasing system size, and for large $V$, the data points generally fall on a line of a power-law behavior $\sim V^{d_f^{*\infty}-d_f^{*d}}$, with $d_f^*$ values listed in Table~\ref{t1}. This confirms that EP in $d=2$ dimension has a larger volume fractal dimension $d_f^*$ than those in $d>2$.

Furthermore, it can be observed that the growth of $C_1^{\infty}/C_1^{d}$ slows down rapidly as $d$ approaches $6$. However, due to sparse data points and significant finite-size corrections, we cannot conclusively determine whether the slow growth behavior of $C_1^{\infty}/C_1^{d}$ for $d=6$ is actually a power-law behavior resulting from different fractal dimensions in $d=\infty$ and $d=6$, or a logarithmic correction at the upper critical dimension (if $d_u=6$). Nevertheless, the numerical results suggest that EP should have an upper critical dimension $d_u\geq6$.

\section{Conclusions} \label{sec-con}

In this study, we investigate the critical behaviors of EP in finite dimensions using an event-based ensemble approach. Through extensive simulations across dimensions $d=2$ to $6$, we find that the scaling behaviors of EP near the realization-dependent dynamic pseudocritical point $\mathcal{T}_L$ are well described by the standard FSS theory in all dimensions. However, in the conventional ensemble, near the critical point $T_c$, the FSS behaviors appear anomalous due to the mixing of critical, subcritical, and supercritical phases, complicating the application of standard FSS theory.

This ensemble inequivalence is intrinsically linked to the FSS behavior of the dynamic pseudocritical point $\mathcal{T}_L$. Based on the specific bond-insertion rules in EP, we propose an argument that the probability distribution of $\mathcal{T}_L$ obeys the central-limit theorem, following a normal distribution. Consequently, the fluctuation of $\mathcal{T}_L$ scales as $V^{-\theta}$, with a dimension-independent exponent $\theta=1/2$. Conversely, the convergence of its mean $T_L\equiv\langle\mathcal{T}_L\rangle$ to the critical point $T_c$ is governed by the correlation-length exponent, described by $T_L-T_c\sim L^{-1/\nu}\sim V^{-1/\nu^*}$. In the conventional ensemble, the coexistence of different exponents $\theta$ and $1/\nu^*$ leads to anomalous FSS behaviors, violating the fundamental assumption that the correlation length is the sole relevant scale for critical phenomena. In contrast, the event-based ensemble samples observables at and near $\mathcal{T}_L$, naturally excluding the effect of $\theta$, thereby allowing standard FSS theory to be applied effectively, enabling accurate determination of critical exponents.

Our precise determination of percolation thresholds and critical exponents across various dimensions reveals a distinct dimension-dependence compared to standard bond percolation, highlighting variations in critical behaviors with dimensionality. Notably, we observed differences in critical exponents between EPs governed by product and sum rules, suggesting that the universality of EP is sensitive to the specific bond-insertion rule.

While our findings provide a robust foundation for understanding EP, several intriguing questions remain. First, it is important to further explore the distinct dimension dependence of EP, including an exact determination of the upper critical dimension. Second, although we have shown that EPs governed by product and sum rules belong to different universality classes, it remains an open question whether EPs under some of the typical bond-insertion rules~\cite{Boccaletti2016} share the same universality. Third, beyond EP, our event-based method has also proven effective in accurately extracting the critical behaviors of high-dimensional percolation~\cite{Li2024}. It would be interesting to investigate whether this method can similarly capture the critical behaviors of discontinuous or hybrid percolation transitions, such as $k$-core percolation~\cite{Dorogovtsev2006,Shang2020,Gao2024}, and percolation on interdependent networks~\cite{Baxter2012,Li2020,Gross2022}.

\section*{Acknowledgments}

The research was supported by the National Natural Science Foundation of China under Grant No.~12275263, the Innovation Program for Quantum Science and Technology under Grant No.~2021ZD0301900, and Natural Science Foundation of Fujian province of China under Grant No.~2023J02032. The research of M.L. was also supported by the Fundamental Research Funds for the Central Universities (No.~JZ2023HGTB0220).

\bibliography{ref}

\end{document}